\documentclass[iop]{emulateapj}
\usepackage{amsmath}

\usepackage{hyperref}
\hypersetup{
    colorlinks=true,
    linkcolor=blue,
    citecolor=black,
    filecolor=magenta,      
    urlcolor=cyan,
}

\newcommand{\reffigure}[1]{Figure~\ref{Figure #1}}

\def\lapprox{\hbox{\lower .8ex\hbox{$\,\buildrel < \over\sim\,$}}}
\def\gapprox{\hbox{\lower .8ex\hbox{$\,\buildrel > \over\sim\,$}}}

\begin{document}

\title{Gamma--rays from kilonovae and the cosmic gamma--ray background}

\author{Pilar Ruiz--Lapuente}
\affiliation{Instituto de F\'{\i}sica Fundamental, Consejo Superior de
Investigaciones Cient\'{\i}ficas, c/. Serrano 121, E--28006, Madrid, Spain}
\affiliation{Institut de Ci\`encies del Cosmos (UB--IEEC),  c/. Mart\'{\i}
i Franqu\'es 1, E--08028, Barcelona, Spain}

\author{Oleg Korobkin}
\affiliation{Center for Theoretical Astrophysics, 
             Los Alamos National Laboratory, 
             Los Alamos, NM, 87545, USA}
\affiliation{Joint Institute for Nuclear Astrophysics - 
             Center for the Evolution of the Elements, USA}
\affiliation{Computer, Computational, and Statistical Sciences Division, 
             Los Alamos National Laboratory, Los Alamos, NM, 87545, USA}


\begin{abstract}

\noindent
The recent detection of the gravitational wave event GW170817, produced by the 
coalescence of two neutron stars, and of its optical--infrared counterpart, 
powered by the radioactive decay of r--process elements, has opened a new 
window to gamma--ray astronomy: the direct detection of photons coming from 
such decays. Here we calculate the contribution of kilonovae to the diffuse 
gamma--ray background in the MeV range, using recent results on the spectra of 
the gamma--rays emitted in individual events, and we compare it with that from 
other sources. We find that the contribution from kilonovae is not dominant in
such energy range, but within current uncertainties, its addition to 
other sources might help to fit the observational data.    

\end{abstract}


\section{Introduction}
Up to this day, there is no clear understanding of which  sources 
or emission mechanisms may account for the MeV background 
\citep{deangelis18,ajello19,tatischeff19,McEnery19}. On the low--energy side, 
from X--ray 
energies up to around 0.3 MeV, AGNs and 
Seyfert galaxies provide most of the emission \citep{madau94,ueda03}, but these 
contributions sharply cut off at $E\gapprox0.3$~MeV. At energies in the 50 MeV 
to the GeV range, blazars, star--forming galaxies and radio galaxies can 
explain the observed background \citep{ajello15, dimauro15}.

A potential candidate for the missing source of MeV photons could be associated 
with the sites of heavy-element nucleosynthesis. However, the sites of the 
nucleosynthesis of heavy elements produced by rapid sequences of neutron 
captures, the r--process elements, has long been a matter of debate. The 
innermost parts of the ejecta from gravitational--collapse supernovae 
\citep{woosley94}, material expelled in the coalescence of two neutron stars
(NSs) or of a NS and a black hole (BH) \citep{lattimer74}, and jets from 
magnetorotationally-driven supernovae \citep{winteler12} have been proposed. In 
this study, we focus on the neutron star mergers and the gamma emission they 
can produce.

\cite{li98} were first to propose that the merger of two NSs should produce an
electromagnetic transient, later named ``macronova'' \citep{kulkarni05} or 
``kilonova'' (KN being the currently most used term), powered by the 
radioactive decay of the merger debris and seen in optical and infrared bands 
\citep{metzger10a}. Such debris would be rich in r--process elements and their 
radioactivity would heat the ejected material and make it luminous. Subsequent
studies of the heating rates \citep{metzger12, korobkin12, lippuner15} found 
that the late--time bolometric light curve of the KN would provide evidence of 
the radioactive material and enable to estimate the amount of r--process 
elements produced in the merger. A re--analysis of the afteglow light curves of 
nearby short gamma--ray bursts (sGRBs) by \cite{jin16}, now generally 
attributed to the merger of two NSs \citep{eichler89,nakar07,berger14}, 
suggests that KNe are always present in the afterglows of the sGRBs 
\citep[see also][]{tanvir13,kasliwal17}.

The detection of GW170817, the gravitational wave signal of a binary NS 
inspiral \citep{abbott17} was followed by the discovery of an electromagnetic 
counterpart, a KN \citep{arcavi17, coulter17, lipunov17, pian17,soaressantos17, 
tanvir17, valenti17}, known as DLT17ck (and also as SSS17a and as AT 2017gfo). 
A sGRB (GRB 170817A), consistent with the gravitational wave signal location, 
was also detected, two seconds later, by the Gamma--Ray Burst Monitor aboard 
the {\it Fermi} spacecraft \citep{goldstein17}.  A second event, GW190425,
 also corresponds to the merger of two NSs 
(Abbott et al. 2020). A sGRB, GRB 190425, was detected with the gamma--ray 
spectrometer SPI aboard the {\it INTEGRAL} 
observatory (Pozanenko et al. 2019), but no observations are available at 
other wavelengths.

\bigskip

\noindent
The observations of the KN associated with GW170817 show that, as predicted, 
it originated from neutron--rich matter unbound from the 
system \citep{mccully17,smartt17,rosswog18}. Two distinct components of the 
KN were
clearly identified: an early blue KN, peaking in optical bands \citep{evans17},
 and a 
late, infrared KN \citep{tanvir17}. The blue peak would be produced 
by a disk--driven wind enriched with lighter r--process elements (Kasliwal 
et al. 2017), while the more slowly evolving infrared emission would be 
powered by the decay of the lanthanide--rich material, dynamicallly ejected
at the merger. This is in agreement with the theory, which predicts that the 
dynamical ejecta from mergers will produce lanthanide-rich composition and 
peak in the infrared, while secondary postmerger outflows will result in a less 
neutron-rich composition, leading to lighter r--process without lanthanides 
\citep{kasen13,barnes13,grossman14}. In this study, we employ a similar 
two--component model for gamma--ray emission: very neutron--rich 
``dynamical ejecta'' and ``wind'', their emission lasting for 
about one month (further details in Section~\ref{sec:rprocess}).

Gamma--rays are emitted by the ejecta at all epochs. They fall, as in the case 
of Type Ia supernovae (SNe Ia), within the MeV range, precisely a region of the 
cosmic gamma--ray background spectrum where there are no known sources that 
satisfactorily fit the observations \citep{ajello19}. For the production of 
gamma--rays, here we consider both the ``kilonova'' phase lasting for about a 
month, 
and a ``remnant'' phase, 
due to long-lived residual nuclides from the r-process, and
extending up to {$\sim$ 10$^{6}$} years 
\citep[see][]{korobkin19}.

In previous studies, \cite{hotokezaka16} pioneered detailed computation of the
gamma-ray spectra from kilonovae. They used a line-broadening approach to 
simulate
the ejecta expansion with subrelativistic speeds (appropriate for epochs earlier
than about one week since the merger). \cite{li19} extended the calculation of 
the
spectrum with a semi-analytic model for radiative transport and nuclear decay
chains. In \cite{korobkin19}, we used a full 3D Monte Carlo radiative transport
code \citep{hungerford03, hungerford05} to simulate the emission in the kilonova
phase, and a simple line broadening to extend the emission spectra to the 
remnant
state (over 100~kyr). In this study, we apply the computed spectra at both
phases to derive the contribution of KNe to the diffuse gamma--ray background.
We will see that KNe do not appear to give a dominant contribution to the
background but, within current uncertainties, might, together with that from
SNe~Ia, improve the fit to the observational data when added to another,
dominant source.

The paper is organized as follows: first we deal with the KN rates in
Section~\ref{sec:knrates} and in Section~\ref{sec:rprocess} with the r--elements
yields. The observations of the gamma--ray background in the MeV range are
presented in Section~\ref{sec:background}. The input gamma--ray spectra and the
method used to calculate the KN background are described in
Section~\ref{sec:modeling}. The results are presented and discussed in
Section~\ref{sec:results}. Section~\ref{sec:summary} summarizes the present
study, gives its conclusions and points to ways for improvements.

\section{The kilonova rates}
\label{sec:knrates}

The rates of KN engines --- neutron star mergers --- have been estimated
theoretically in multiple studies \citep[e.g.][]{kalogera04a, kalogera04b,
kim15, wanderman15}.  \cite{abbott20}, from the detections of GW170817 
and GW190425, infer
a rate of {1090$^{+1720}_{-800}$ Gpc$^{-3}$ yr$^{-1}$}. For that rate, as we
will see, the kilonova contribution to the gamma--ray background would be very
minor. But, as noted by \cite{dellavalle18}, the rate of NS--NS mergers is still
very poorly constrained. Recently, \cite{yang17}, based on the light curve of AT
2017gfo/DLT17ck (the electromagnetic counterpart of GW170817) and on the results
of the DLT Supernova search, set an upper limit of {$0.99\times10^{-4}\
^{+0.19}_{-0.15}$ Mpc$^{-3}$ yr$^{-1}$} to the local rate ({$d < 40$~Mpc}) of
binary NS mergers. We will use it as our reference rate, but we equally
 consider the upper and lower limit given by \cite{abbott20}.

For calculating the contribution of KNe to the cosmic gamma--ray background, we
need to estimate how the rates have varied along $z$. Massive stars promptly 
become neutron stars, but there can be a considerable time interval between the 
formation of binary systems made of two NSs and the merger of the two objects. 
\cite{wanderman15} find that there is a delay of {3--4 Gyr} of the mergers 
relative to the global star formation rate. 

For the cosmic star formation rate, we use the results of \cite{cucciati12}. We 
derive the binary neutron star coalescence rate assuming an average delay time 
of 3.5 Gyr and we normalize the rates to the upper limit set by \cite{yang17}
to the local rate. The resulting KN rate, $R_{KN}(z)$ (Mpc$^{-3}$ yr$^{-1}$), 
is shown in \reffigure{1}.

\begin{figure}[ht!]
\centering
\includegraphics[width=1.0\columnwidth]{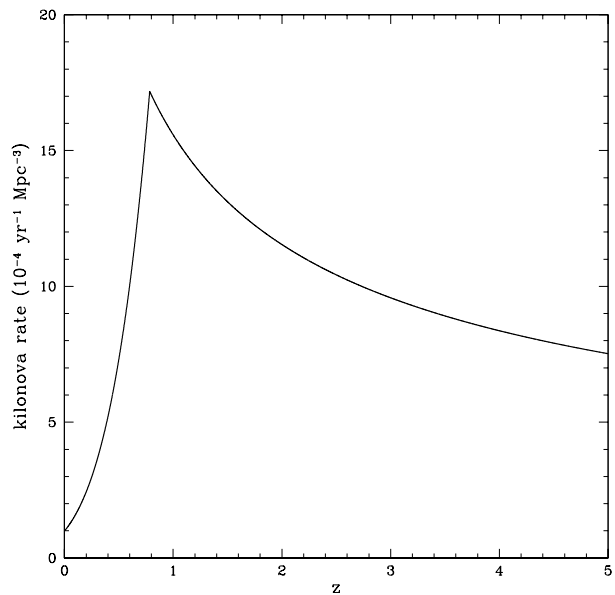}
\caption{The KN rate as a function of redshift $z$, for an average 
time delay of 3.5 Gyr between star formation and NS--NS mergers, based 
on the cosmic star formation rate of \cite{cucciati12}, normalized 
to the upper limit to the local KN rate set by \cite{yang17}.}  
\label{Figure 1}
\end{figure}

\section{The r--process yields from kilonovae}
\label{sec:rprocess}

As pointed out in the Introduction, NS--NS  mergers are main candidates to the 
production of the r--process elements. Their relevance in this respect depends 
on the rate of the mergers and on the amount of r--process elements produced in 
each event. We have dealt with the first point in the previous Section.

Concerning the second point, the amount and composition of the ejecta depend on
how and when during the merger the material has become unbound \citep[for a
review, see][]{metzger19}. The most neutron-rich component known as ``dynamical
ejecta'' gets unbound by dynamical tide and attains ideal conditions for robust
r--process nucleosynthesis \citep{freiburghaus99, korobkin12, bauswein13}. Other
types of ejecta include the material ``squeezed'' from the binary contact
interface \citep{goriely11a, wanajo14}, neutrino-driven outflows from the merged
hypermassive neutron star \citep{dessart09a, perego14a}, from the accretion disk
\citep{fernandez15, miller19}, or fallback material \citep{desai19}. These types
may produce full r--process but in general are not expected to produce robust,
main r--process nucleosynthesis due to their lack of mechanism to ensure high
neutron richness, necessary for fission cycling \citep{holmbeck19}.

Observationally, ejecta can be classified into two types: lanthanide-rich ejecta
producing the late ``(infra-)red'' KN, and lanthanide-poor component,
responsible for the ``blue'' KN~\citep{metzger19}. Below, we will refer to these
two components as ``dynamical ejecta'' and ``wind'', and use representative
compositions described in \cite{korobkin19} (see their Fig.~3 and Table~1).
Regarding the masses, theoretical estimates vary around $10^{-2}$~M$_\odot$
within about two orders of magnitude \citep{hotokezaka13, bauswein13, radice16,
dietrich17}. Theoretical models for GW170817 give a 
similar range of estimates for
the two components, as shown in Table~1 of \cite{cote18} \citep[see also][for
more complete summary]{ji19}. Somewhat higher values were deduced for previous
afterglow excess events \citep{tanvir13, piran14, jin16, wollaeger18}, although
here a selection effect might play a role.

In this study, we adopt a conservative range from 10$^{-4}$ M$_{\odot}$ to 0.1
M$_{\odot}$ for both ejecta components, to cover the described uncertainties.

\section{The observed diffuse gamma--ray background}
\label{sec:background}

\begin{figure}[ht!]
\centering
\includegraphics[width=1.0\columnwidth]{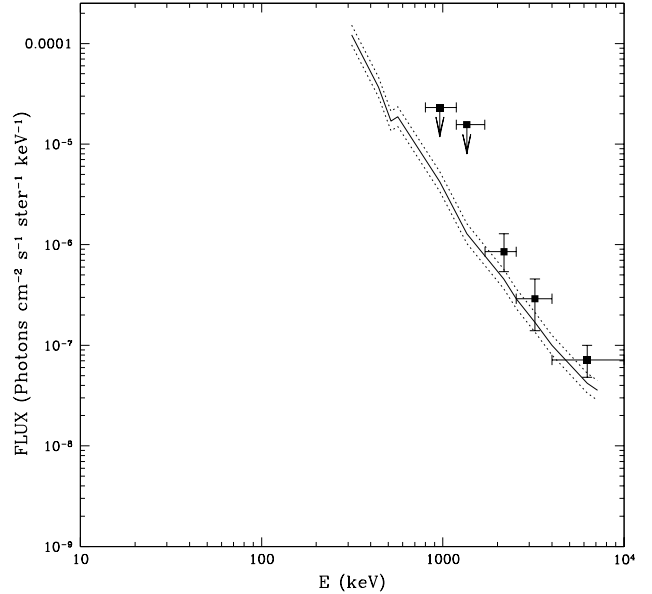}
\caption{The diffuse gamma--ray background in the MeV range, as measured by the
{\it Solar Maximum Mission} \citep[solid line, dotted lines being the
1\,$\sigma$ upper and lower limits:][]{watanabe99} and by COMPTEL \citep[black
squares:][]{kappadath96}.}
\label{Figure 2}
\end{figure}

The present measurements of the gamma--ray background in the MeV range (from 100
keV to 10 MeV) mostly come from two space missions: the {\it Solar Maximum
Mission} \citep[{\it SMM};][]{watanabe99} and COMPTEL \citep{kappadath96,
weidenspointner99, weidenspointer20}. More recent gamma--ray missions, either
have not covered this energy range (like {\it Fermi}) or have not yet produced
available data on the background (like {\it INTEGRAL}). The measurements are
shown in \reffigure{2}. As we see, the slope of the emission spectrum has a
steep decrease with increasing energy, from a few hundred keV to 10 MeV (flux
$\propto E^{-2.5}$, approximately). It changes to a flatter slope around 10 MeV
and beyond. An intense extragalactic source (or the addition of several) is
needed in the MeV window \citep[see, for instance,][]{lacki14}.

As we have seen, the gamma--ray emission from KNe, both in their dynamical and
remnant stages, falls within the above range, hence the interest of modeling
their contribution to the diffuse background emission.

\section{Modeling the kilonova gamma--ray background}
\label{sec:modeling}

To calculate the KN contribution to the diffuse gamma--ray background, we
integrate the evolving luminosity over the time in which the ejecta 
appreciably emit
gamma--rays.  First we will consider the contribution in the KN phase, which
includes the dynamical and wind components This phase lasts for about a month.
Afterwards, we add the contribution from the remnant phase, extending up to
$\sim10^6$ yr. We use the spectra of four models, calculated in
\cite{korobkin19} (see their Fig.~4): 
the evolution of these spectra over 32 days is
reproduced in \reffigure{3}, for their model labeled Ak \citep[see Table 1
in][]{korobkin19}. This model represents a very neutron-rich outflow with,
generating the main (robust) r-process through fission cycling. We have checked
that for the remaining three models As, S1 and S2 the resulting contributions to
the background are very similar. Our standard case corresponds to their adopted
values for the mass ejected: 0.0065 M$_{\odot}$ in the dynamical ejecta
component and 0.03 M$_{\odot}$ in the wind component. Given the wide range of
ejected masses derived from the observations of GW170817 by different authors
(see Section \ref{sec:rprocess} above) we also calculate the KN background for
both the highest and the lowest masses in the range, in order to obtain an idea
of its upper and lower limits. We are aware of the roughness of our treatment,
since scaling the emission by the mass of the ejecta neglects the change in
material opacity to the gamma--rays in KN phase, and a full calculation of the
spectra should instead be made for each case.  However, the effect of a finite
opacity is significant only at the initial epoch of a day or so, and should be
subdominant to the range of mass uncertainty.

The spectra for the remnant phase of the model Ak are shown in \reffigure{4}.
Here the ejecta is practically transparent and no transport calculation is
needed. To obtain the spectra at different times, we computed the
detailed nuclear
gamma--ray source using Eq.~(1) in \cite{korobkin19} and applied Doppler
broadening with typical remnant expansion velocities for each epoch \citep[see
Fig.~6 in][]{korobkin19}. The \reffigure{4} also compares the 
Doppler-broadened spectrum
with the spectrum from the full radiative transfer model (dashed line) at $t =
32$~days. Two spectra show an excellent agreement, demonstrating the validity of
our approach for the remnant phase.

\begin{figure}[ht!] 
\centering
\includegraphics[width=1.0\columnwidth]{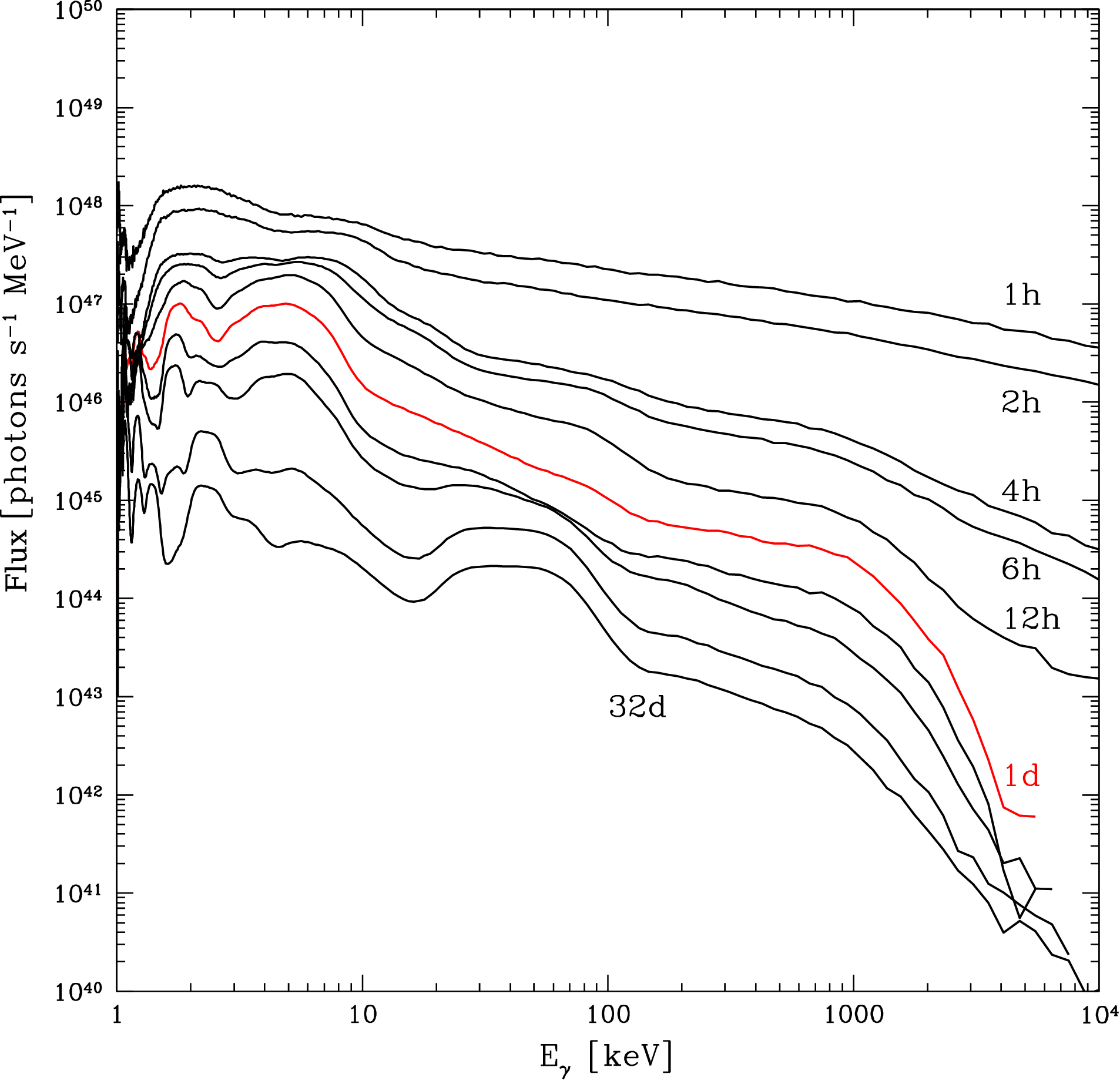}
\caption{Evolution of the gamma--ray spectra in the KN phase, up to 32~days
after a NS--NS merger, computed with 3D radiative transfer code for the
neutron-rich dynamical ejecta model Ak \citep[from][see their Table
1]{korobkin19}.}
\label{Figure 3}
\end{figure}

\begin{figure}[ht!] 
\centering
\includegraphics[width=1.0\columnwidth]{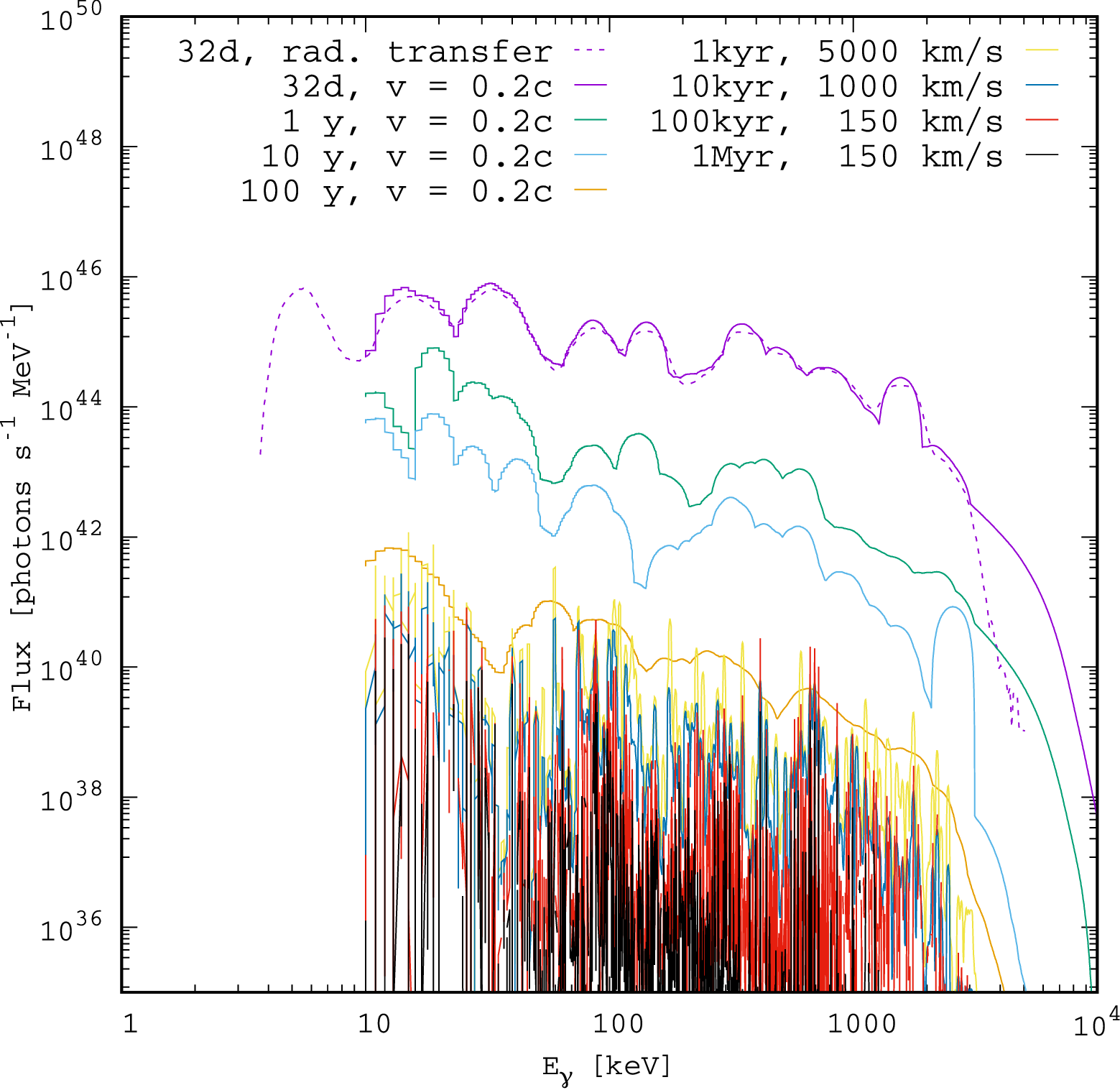}
\caption{Evolution of the gamma--ray spectra in the remnant phase, from 32 days
up to 10$^{6}$~yr, for the neutron-rich dynamical ejecta model Ak. For each
solid line, the detailed isotopic decay spectrum is Doppler-broadened with the
remnant expansion velocity typical for the remnant phase at this epoch (shown in
the legend). The dashed line represents the full radiative transfer model Ak at
$t=32$~days, in good agreement with the Doppler-broadened spectrum.} 
\label{Figure 4}
\end{figure}

To model the contribution of KN to the cosmic gamma-ray background, we follow
the same steps as in \cite{ruiz-lapuente16} when modeling that from SNe~Ia.
First, from the gamma--ray emission at different stages in the evolution of the
ejecta, we infer the total number of photons emitted ({photons keV$^{-1}$}) in
each event, and dividing by its duration, $\tau_{KN}$ (taken as 32 days for the
KN phase and as $10^6$~yr for the remnant phase), the average luminosity,
$l_{\gamma}(E)$ ({photons s$^{-1}$ keV$^{-1}$}) is obtained. The number of KNe
per unit of comoving volume, $R'_{KN}(z)$, active at any time, is that of the KN
produced during the previous time interval of duration $\tau_{KN}$, so:
${R'_{KN}(z) = {\rm const} \times R_{KN}(z)}$, the latter being the comoving KN
rate ({KN yr$^{-1}$ Mpc$^{-3}$}) and, for the KN phase, ${\rm const} = 32/365.25
= 0.0876$, while for the remnant phase ${\rm const} = 10^{6}$. The contribution
to the gamma--ray background of the shell at comoving radius $r$ and with
thickness $dr$ is:

\begin{equation}  
dL_{\gamma}(E, z) = 4\pi R'_{KN}(z)l_{\gamma}(E)dV(z)
\end{equation}

\noindent
where

\begin{equation}
dV(z) = D_M^{2}(z)dD_M
\end{equation}

\noindent
$D_M$ being the proper motion distance (in Mpc). The flux received from that
shell ({photons cm$^{-2}$ s$^{-1}$ keV$^{-1}$}) will be:

\begin{equation}
dF_{\gamma}(E, z) = {\frac{1}{4\pi} D_L(z)^{2}}dL_{\gamma}[(z + 1)E, z]
\end{equation}

\noindent
$D_L$ being the luminosity distance (cm). The factor ${(z + 1)}$, multiplying 
$E$, accounts for the redshift of the photons. Then we have:

\begin{equation}
dF_{\gamma}(E, z) = R'_{KN}(z)l_{\gamma}[(z + 1)E]{D_M^{2}(z)\over 
D_L^{2}(z)}dD_M 
\end{equation}

\noindent
Due to time dilation, there should be a factor ${(1 + z)^{-1}}$ multiplying 
the comoving KN rate, but it is canceled by the ${(z + 1)}$ factor accounting
for compression of the energy bins. Since ${D_L = (1 + z)D_M}$, we have

\begin{equation}
dF_{\gamma}(E, z) = {1\over (1 + z)^{2}}R'_{KN}(z)l_{\gamma}[(z + 1)E]dD_M 
\end{equation}

\noindent  
$dD_M$ depends on the cosmological parameters $H_{0}$, $\Omega_{M}$ and 
$\Omega_{\Lambda}$, so we finally have:

\begin{equation}
\begin{split}
F_{\gamma}(E) = {c\over H_{0}} \int_{0}^{z_{\rm lim}}{{1\over (1 + z)^{2}}R'_{KN}
l_{\gamma}[(z + 1)E]} \\ \times e(z, \Omega_{M}, \Omega_{\Lambda})dz
\end{split}
\end{equation}

\noindent
We adopt {$H_{0}$ = 67 km s$^{-1}$ Mpc$^{-1}$}, ${\Omega_{M} = 0.31}$ and 
${\Omega_{\Lambda} = 0.69}$, from the \cite{planck18}, 
assuming a flat universe. The last term in the previous equation is:

\begin{equation}
e(z, \Omega_{M}, \Omega_{\Lambda}) = [(1 + z)^{2}(1 + \Omega_{M}z - 
z(2 + z)\Omega_{\Lambda}]^{-1/2}  
\end{equation}   

\noindent
In order to compare the calculated fluxes with observations, we must divide the
$F_{\gamma}(E)$ above by 4$\pi$, to convert to the units used in reporting 
observed fluxes (photons cm$^{-2}$ s$^{-1}$ keV$^{-1}$ sr$^{-1}$).

\section{Results and discussion}
\label{sec:results}

In \reffigure{5}, the contributions of mergers in their KN phase to the
diffuse gamma--ray background in the 10 keV -- 10 MeV range are compared with
the available observations. The continuous black line corresponds to our
``standard'' case, and the two dotted red lines to our adopted upper and lower
limits for the ejected mass. The two blue, slashed lines correspond to the upper
and lower limits to the local merger rate set by \cite{abbott20}, for the
``standard'' ejecta masses, while the red dot--dashed line is for the
combination of their upper limit on the merger rate with that on the ejected
mass.  We see that, for energies $E_{\gamma}$ between roughly 200~keV and 3~MeV,
the slope of the contribution follows that of the data, but at significantly
lower values of the fluxes, even for our ``standard'' upper limit, than the
observed ones. If we compare with the contribution from SNe~Ia (which is made in
\reffigure{7}), we have that KNe have a higher average luminosity $l_{\gamma}$
than SNe~Ia, but that is countered by the fact of the shorter duration of the
KNe in this phase (32 days against 600 days) and by their lower rate of
occurrence.

\begin{figure}[ht!]
\centering
\includegraphics[width=1.0\columnwidth]{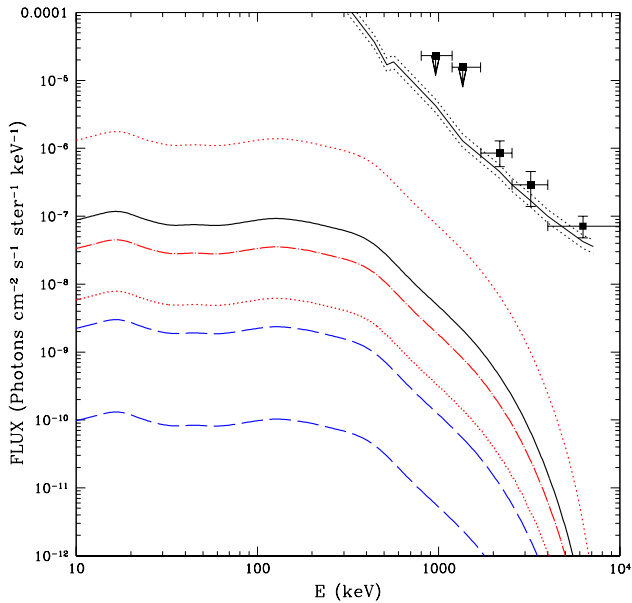}
\caption{The contributions of NS--NS mergers to the cosmic gamma--ray
background in the MeV range, due to the emission of the ejecta in KN phase,
calculated from the spectra in \reffigure{3}, for the standard ejecta masses
(black continuous line) and for the maximum (${0.1\,M_\odot}$) and minimum
(${10^{-4}\,M_\odot}$) masses in the range (red dotted lines).  They are
compared with the observations (continuous black line: {\it Solar Maximum
Mission}, from \cite{watanabe99}, dotted lines being the 1\,$\sigma$ upper and
lower limits; black squares: COMPTEL, from \cite{kappadath96}). The two blue,
dashed lines correspond to the upper and lower limits to the local NS--NS
merger rate set by  \cite{abbott20}, and the red dot--dashed line to the
combination of their upper limit on the KN rate with that on the ejected mass.}
\label{Figure 5}
\end{figure}

In \reffigure{6}, we plot the contribution to the background by NS--NS merger
ejecta in their remnant phase, also the ``standard'' one (continuous red line)
and upper and lower limits estimated as above. The average luminosity $l_\gamma$
is lower than in the KN phase, that being partially compensated by the long
duration of the phase. The contributions calculated for the upper and lower
limits to the local KN rate from  \cite{abbott20} are indicated by the blue
dashed lines, while the dot--dashed one corresponds to the combination of their
upper limit on the merger rate with that on the ejected mass of radioactive
nuclei.

\begin{figure}[ht!]
\centering
\includegraphics[width=1.0\columnwidth]{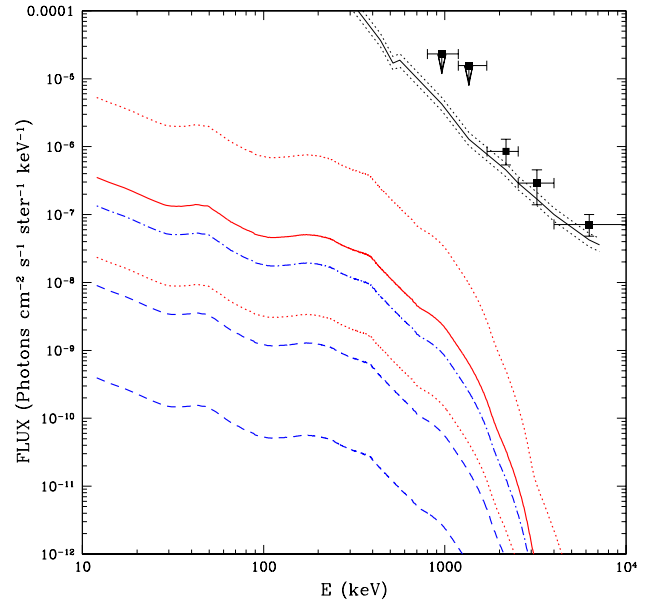}
\caption{Similar to \reffigure{5}, with the contribution of NS--NS mergers
 due to the emission of the ejecta in their remnant phase (red
continuous line) and its upper and lower limits (red dotted lines), calculated
from the spectra in \reffigure{4}. Also shown, by the blue dashed lines, are
the contributions calculated for the upper and lower limits to the local NS--NS
merger rate from  \cite{abbott20}. As in the previous Figure, the 
dot--dashed
line corresponds to the combination of their upper limit on the KN rate with
that for the mass of the ejecta.}
\label{Figure 6}
\end{figure}

In \reffigure{7}, the KN backgrounds (``standard'' and upper and lower limits)
for all phases are compared with that from SNe~Ia \citep[dashed line, adopted
from][]{ruiz-lapuente16}. We also show the result (dot--dashed blue line in the
Figure) of adding our upper limit to the KN contribution to that of the 
SNe~Ia. Equally shown, again, are the backgrounds
corresponding to the upper and lower limits to the local KN rate from
\cite{abbott20} and the combination of that upper limit with the highest
estimate of the mass of the KN ejecta. The coincidence of the slope of the KN
contribution with that of the measured background is remarkable, although it
remains significantly below even at its upper limit.

\begin{figure}[ht!]
\centering
\includegraphics[width=1.0\columnwidth]{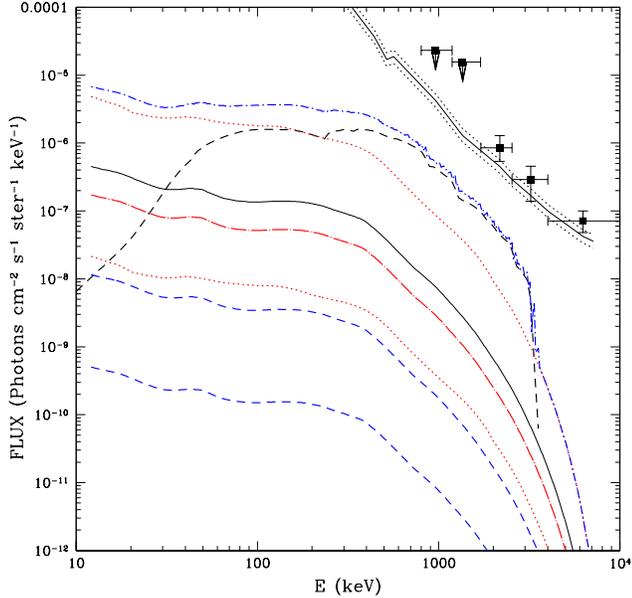}
\caption{The black
 line shows the addition of the contributions of NS--NS mergers 
in the KN and the remnant phase ( \reffigure{4} and \reffigure{5}).
 The red--dotted lines corresponds to the upper and lower limits to the 
ejecta masses in that case.  They are
compared with that of SNe~Ia (black dashed line), taken from
\cite{ruiz-lapuente16}. Also shown (blue dot--dashed line) is the result of
adding the KN contribution at its upper limit to the SN~Ia contribution.
Equally shown (blue dashed lines) are the contributions calculated for the upper
and lower limits to the local KN rates of  \cite{abbott20} and the 
combination
of that upper limit with our highest estimate of the mass of the KN ejecta (red
dot--long dashed line).}
\label{Figure 7}
\end{figure}

There has been another recent calculation of the gamma--ray emission from KNe
\citep{li19}, where a semianalytic model of radiative transfer was introduced.
In order to check the differences arising from two different treatments of the
emission processes, we have also calculated the background using now the spectra
shown in Fig.~14 of \cite{li19} and corresponding to his standard case. The
result is displayed in \reffigure{8}, where it is compared with our own results
for the emission in the KN phase of the ejecta (also in the standard case). We
see that there is some difference, but not a very significant one.

\begin{figure}[ht!]
\centering
\includegraphics[width=1.0\columnwidth]{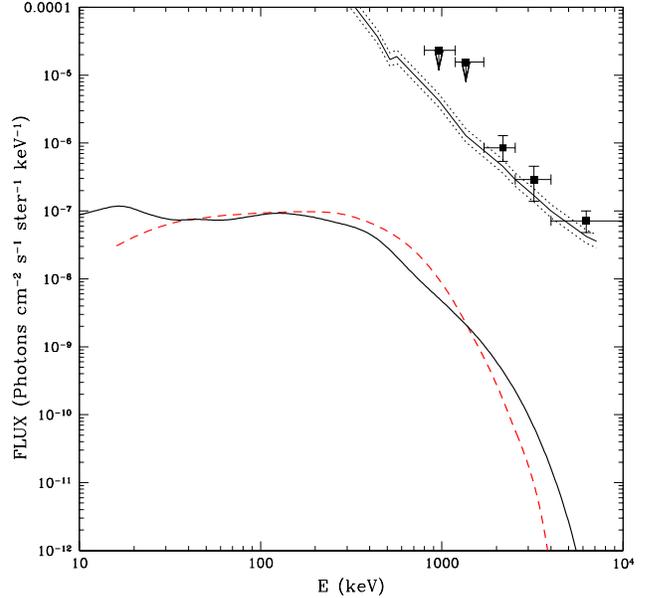}
\caption{Comparison of the contribution of NS--NS mergers to the cosmic
gamma--ray background, in the KN phase, calculated from the emission spectra of
\cite{li19} (red slashed line), with that from the present work (continuous
black line). The background observations are also shown, as in the previous
Figures.}
\label{Figure 8}
\end{figure}

A further comparison is made in \reffigure{9}, analogous to \reffigure{7}, of
the gamma--ray background calculated from \cite{li19} and its upper and lower
limits (estimated as in our case), with the contribution from SNe~Ia. Also shown
is the result of adding the KN contribution, taken at its upper limit, in this
case, to the SNe~Ia one.

\begin{figure}[ht!]
\centering
\includegraphics[width=1.0\columnwidth]{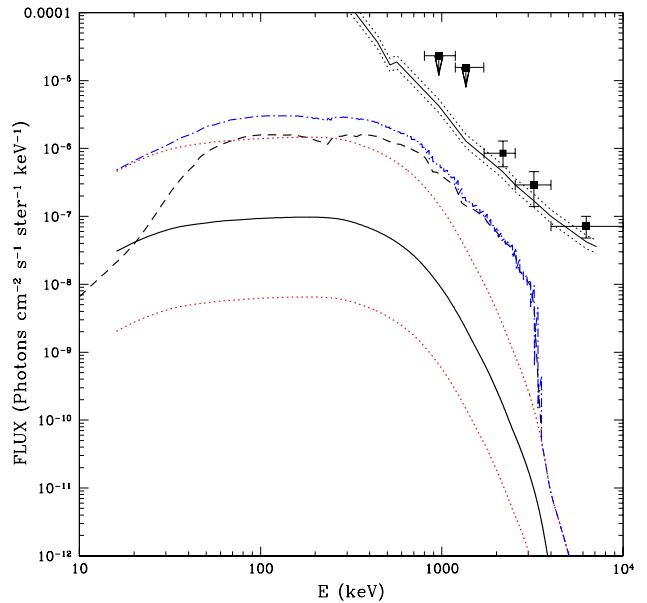}
\caption{Analogous to \reffigure{7}, for the contribution of KNe to the cosmic
gamma--ray background obtained from \cite{li19} (continuous black line: standard
case; dotted red lines, upper and lower limits). Also shown (blue dot--dashed
line) is the result from adding together the upper limit of the KN contribution
and that from SN~Ia (shown by the black dashed line).}
\label{Figure 9}
\end{figure}   

We must stress that our ``reference'' results shown in Figures~\ref{Figure
5}--\ref{Figure 7} (and also the comparisons in Figures~\ref{Figure
8}--\ref{Figure 9}) correspond to the normalization of the KN rates to the upper
limit to the local rate ($d <$ 40 Mpc) set by \cite{yang17}, which is
significantly higher than the upper limit set by  \cite{abbott20} from the
observations of GW170827 and GW190425, but still well within the 
uncertainty factor for the NS--NS merger rate indicated by 
\cite{dellavalle18}. In any case, as soon as the local merger rate
estimates are refined, the Figures above need only to be rescaled to the new
value of the rate.  However, it seems improbable to reach the MeV background for
any viable rate.

From the results obtained, we see that the KN contribution to the cosmic
gamma--ray background (subjected to the previous caveat), although being
non--negligible, appears minor as compared with that of SNe~Ia. Only improbably
high rates of NS--NS mergers and/or higher ejecta masses might change the
situation. We must note, however, that even the contribution from SNe~Ia falls
short of explaining the observations in the MeV range, and some additional
source has to be invoked. Flat--spectrum radio quasars \citep{ajello09} (with a
particular assumption about the location of the inverse--Compton peak in their
spectra, due to the same electron population that emits the synchrotron bump),
could explain the bulk of the background, though its slope in the MeV region
does not really correspond to that of the data \citep[see Figs. 15 and 16
of][]{ajello09}. 
  Hidden cores of AGN could 
also originate MeV  radiation. In that case,
 the MeV emission should correlate with 
the TeV neutrino background \citep{murase19}. This contribution to the 
gamma--ray background still needs to be fully calculated.
The SN~Ia contribution is quite 
 substantial
\citep[see Fig. 13 in][]{ruiz-lapuente16}.
We see, in Figures~\ref{Figure 5}--
\ref{Figure 8}, that a KN contribution close to our upper limit or above would
act analogously.

The question arises of the upper limit set by the observed abundances of
r--process elements to the combination of NS--NS merger rate and r--process
element yields in each merger event  \citep{vangioni16}.
 \cite{bovard17} have addressed this point
(see their Fig.~14). They adopt a simple model of Galactic evolution \citep[see
also][]{cote17}, where the yields from each event accumulate along the history
of the Galaxy. It must be taken into account, however, that the KN ejecta move
at velocities ${v_{dyn} \sim 0.2c}$ (dynamical component) and ${v_{wind} \sim
0.08c}$ (wind component). Such velocities are much higher that those of the SN
ejecta and can allow escape from the Galaxy of a large fraction of the material,
especially given that NS--NS mergers, as shown by the distribution of SGRBs in
other galaxies \citep[see][for instance]{fong13}, occur far from the regions of
star formation and even far away from the bulk of the stellar mass.  That is due
to the two successive kicks accompanying the formation of the two NSs \citep[but
see][]{piran05b,beniamini16}. Given the very low densities of interstellar
matter there, the ejecta should almost be expanding in a void and that, together
with their high velocities, would made them leave the Galaxy (that would in no
way affect their contribution to the background, though). Therefore, no robust
upper limit can easily be derived to the combination of KN rate with the amount
of ejected material per NS--NS merger.

Another point to be considered is that the gamma--ray emission from the
radioactive elements created by the NS--NS mergers has a cut--off at {$\sim$20
MeV}. Although the location of the cutoff is much smeared when adding the
contributions to the gamma--ray background at different redshifts, some feature
might still show if the emission from KNe were dominant for {$E_{\gamma} <$ 20
MeV}. But the slope of the observed spectrum just flattens at {$E \simeq$ 10
MeV} (when going to higher energies), so not much can be concluded.

\section{Summary and conclusions}
\label{sec:summary}

Neutron star merger ejecta are emitters of gamma--rays in the MeV range, both in
the kilonova (KN) phase of the ejection of r--process rich material and in the
much longer remnant phase of the expansion. Based on recent calculations of the
spectra of those emissions \citep{korobkin19, li19}, we have estimated the
contribution of NS--NS mergers to the cosmic gamma--ray background, which is
significant in the range from 10~keV to a few MeV, just where there is a gap
with no clear source (or sources) that can explain the observational data. We
take into account the current, considerable discrepancies about the amount of
material ejected in the mergers and also on the cosmic rates of these events.

We find that, within the current upper limits, the contribution of mergers falls
short to explain the observed background, although its slope coincides with that
of the observational data.  At its upper limit, added to the contribution from
SNe~Ia \citep{ruiz-lapuente16} and from flat--spectrum radio quasars
\citep{ajello09}, it would help to fit the data in the MeV region.

We have not included the contribution to the background from NS--BH mergers, for
which there are no available calculations of the emitted gamma--ray spectra.  In
any case, their occurrence rate should be lower than that of NS--NS mergers,
although the ejected masses might be larger.

Upcoming detections of gravitational--wave events and the observation of their 
electromagnetic counterparts should clarify the local merger rate 
and the amounts of r--process rich material ejected in the mergers. Eventually, 
the emitted gamma--rays from nearby events should be observed by some of the 
space missions planned for the near future. In particular, the {\it enhanced 
ASTROGAM} ({\it e--ASTROGAM}) space mission \citep{tatischeff18, deangelis18} 
is conceived to study, with unprecedented sensitivity and resolution, the 
energy range from 300~keV to 3~GeV. Also, the {\it All--Sky--ASTROGAM} 
\citep{tatischeff19}, recently proposed as the ``FAST'' (F) mission of the 
European Space Agency, would explore the range from 100~keV to a few hundred 
MeV, with a very large field of view in this case.   
On the US side, NASA is considering a medium energy gamma--ray mission covering from 200 keV to 10 GeV, the 
 {\it All-sky Medium 
Energy Gamma-ray Observatory, AMEGO} \citep{McEnery19}. Both 
{\it e--ASTROGRAM} and {\it AMEGO} are planned 
to have a high sensitivity to measure the spectral energy distribution (SED) 
both in continuum and  line emission with good spectral resolution and covering
a wide field of view. They are planned to allow polarization studies as well. 
From all 
the above the spectral MeV range would be understood with high observational
precision in the next decade and give an answer to the various 
 candidate contributors. 
  
\acknowledgements
\section*{Acknowledgements}

 We thank Dieter Hartmann for conversations about the planned missions mentioned
in this paper. We are also grateful to Christopher Fryer, Aimee Hungerford and 
 Stephan Rosswog for valuable
input during the preparation of the paper. P.R-L is funded by the Ministry of 
Science and Education of Spain under grant PGC2018--095157--B--100. 
O.K. acknowledges funding from the 
US Department of Energy through the Los Alamos National Laboratory. Los Alamos
National Laboratory is operated by Triad National Security, LLC, for the
National Nuclear Security Administration of U.S. Department of Energy (Contract
No. 89233218CNA000001). Research presented in this article was supported by the
Laboratory Directed Research and Development program of Los Alamos National
Laboratory under projects number 20190021DR and 20200145ER. 
LANL calculations were performed on LANL Institutional Computing resources.


\end{document}